\documentclass{ws-procs975x65}

\begin{document}

\title{SUPERSYMMETRIC $SO(N)$ FROM A PLANCK-SCALE STATISTICAL PICTURE}

\author{ROLAND E. ALLEN}

\address{Department of Physics and Astronomy, Texas A\&M University\\
College Station, Texas 77843, U.S.A.\\
$^*$E-mail: allen@tamu.edu\\
http://www.physics.tamu.edu/}

\begin{abstract}
Several refinements are made in a theory which starts with a Planck-scale
statistical picture and ends with supersymmetry and a coupling of
fundamental fermions and bosons to $SO(N)$ gauge fields. In particular, more
satisfactory treatments are given for (1)~the transformation from the
initial Euclidean form of the path integral for fermionic fields to the
usual Lorentzian form, (2)~the corresponding transformation for bosonic
fields (which is much less straightforward), (3)~the transformation from an
initial primitive supersymmetry to the final standard form (containing,
e.g., scalar sfermions and their auxiliary fields), (4)~the initial
statistical picture, and (5)~the transformation to an action which is
invariant under general coordinate transformations.
\end{abstract}


\bodymatter

\section{Introduction}

This paper contains several refinements of ideas proposed earlier, in the
context of a theory which starts with a statistical picture at the Planck
scale and ultimately results in a supersymmetric $SO(N)$ gauge theory~\cite 
{allen-97,allen-02,allen-03}. The present treatment supersedes previous
versions.

\section{Transformation to Lorentzian path integral: fermions}

We begin with the following low-energy action for the (initially massless)
fundamental fermions and bosons, which follows from essentially the same
arguments as in Refs.~1 and 2, within a Euclidean picture (as in Eq.~(6.17)
of Ref.~1) but with all the components of the vierbein real (as in
Eq.~(3.45) of Ref.~2): 
\begin{eqnarray}
S &=&S_{f}+S_{b} \\
S_{f} &=&\int d^{4}x\,\,\psi _{f}^{\dag }\left( x\right) \,e_{\alpha }^{\mu
}\,i\sigma ^{\alpha }D_{\mu }\,\psi _{f}\left( x\right)   \label{eq2} \\
S_{b} &=&\int d^{4}x\,\,\psi _{b}^{\dag }\left( x\right) \,e_{\alpha }^{\mu
}\,i\sigma ^{\alpha }D_{\mu }\,\psi _{b}\left( x\right)   \label{eq3} \\
D_{\mu } &=&\partial _{\mu }-iA_{\mu }^{i}t_{i}
\end{eqnarray} 
in an obvious notation (which is defined in Refs. 1 and 2). The
transformation of $S_{b}$ to the standard form for scalar bosons will be
treated in the next section, and here we consider $S_{f}$ only.

A key point is that the low-energy \textit{operator} $e_{\alpha }^{\mu
}\,i\sigma ^{\alpha }D_{\mu }$ in $S_{f}$ is automatically in the correct
Lorentzian form, even though the initial \textit{path integral} is in
Euclidean form. It is this fact which permits the following transformation
to a Lorentzian path integral: Within the present theory, neither the fields
nor the operators (nor the meaning of the time coordinate) need to be
modified in performing this transformation.

In a locally inertial coordinate system, the Hermitian operator within $ 
S_{f} $ can be diagonalized to give 
\begin{eqnarray}
S_{f} &=&\int d^{4}x\,\,\psi _{f}^{\dag }\left( x\right) \,i\sigma ^{\mu
}D_{\mu }\,\psi _{f}\left( x\right)  \label{eq5} \\
&=&\sum\nolimits_{s}\,\,\widetilde{\psi }_{f}^{\ast }\left( s\right)
\,a\left( s\right) \,\widetilde{\psi }_{f}\left( s\right)
\end{eqnarray}
where 
\begin{equation}
\psi _{f}\left( x\right) =\sum\limits_{s}U\left( x,s\right) \,\widetilde{
\psi }_{f}\left( s\right) \quad ,\quad \widetilde{\psi }_{f}\left( s\right)
=\int d^{4}x\,U^{\dag }\left( s,x\right) \,\psi _{f}\left( x\right)
\label{eq7}
\end{equation}
with 
\begin{eqnarray}
\,\,i\sigma ^{\mu }D_{\mu }U\left( x,s\right) &=&a\left( s\right) U\left(
x,s\right)  \label{eq8} \\
\int d^{4}x\,U^{\dag }\left( s,x\right) U\left( x,s^{\prime }\right)
&=&\delta _{ss^{\prime }}\;\quad ,\quad \;\sum\limits_{s}U\left( x,s\right)
U^{\dag }\left( s,x^{\prime }\right) =\delta \left( x-x^{\prime }\right)
\end{eqnarray}
so that 
\begin{equation}
\int d^{4}x\,U^{\dag }\left( s,x\right) \,i\sigma ^{\mu }D_{\mu }U\left(
x,s^{\prime }\right) =a\left( s\right) \delta _{ss^{\prime }}\;.
\end{equation}
$U\left( x,s\right) $ is a multicomponent eigenfunction (which could also be
written $U_{s}\left( x\right) $ or $\left\langle s\,|\,x\right\rangle $,
with $U^{\dag }\left( s,x\right) $ written as $U_{s}^{\dag }\left( x\right) $
or $\left\langle x\,|\,s\right\rangle $). Alternatively, $U\left( x,s\right) 
$ is a unitary matrix which transforms $\widetilde{\psi }_{f}\left( s\right) 
$ into $\psi _{f}\left( x\right) $. There is an implicit inner product in 
\begin{eqnarray}
U^{\dag }\left( s,x\right) \,\psi _{f}\left( x\right)
&=&\sum\limits_{r}U_{r}^{\dag }\left( s,x\right) \,\psi _{f,r}\left( x\right)
\\
U^{\dag }\left( s,x\right) U\left( x,s\right)
&=&\sum\limits_{r,a}U_{r,a}^{\dag }\left( s,x\right) U_{r,a}\left( x,s\right)
\end{eqnarray}
with the $2N$ components of $\psi _{f}\left( x\right) $ labeled by $ 
r=1,...,N $ (spanning all components of all gauge representations) and $ 
a=1,2 $ (labeling the components of Weyl spinors), and with $s$ and $x,r$
each formally regarded as having $2N$ values.

Evaluation of the Euclidean path integral (a Gaussian integral with
Grassmann variables) is then trivial for fermions: as usual, 
\begin{eqnarray}
Z_{f} &=&\int \mathcal{D}\,\psi _{f}^{\dag }\left( x\right) \,\mathcal{D} 
\,\psi _{f}\left( x\right) \,\,\,e^{-S_{f}} \\
&=&\prod_{x,ra}\int d\,\psi _{f,ra}^{\ast }\left( x\right) \int d\,\psi
_{f,ra}\left( x\right) \,e^{-S_{f}} \\
&=&\prod_{s}z_{f}\left( s\right) 
\end{eqnarray} 
with 
\begin{eqnarray}
z_{f}\left( s\right)  &=&\int d\,\widetilde{\psi }_{f}^{\ast }\left(
s\right) \,\int d\,\widetilde{\psi }_{f}\left( s\right) \,e^{-\widetilde{ 
\psi }_{f}^{\ast }\,\left( s\right) \,a\left( s\right) \,\widetilde{\psi } 
_{f}\left( s\right) } \\
&=&a\left( s\right) 
\end{eqnarray} 
since the Jacobian $J$ of the transformation in the path integral is unity: 
 \begin{eqnarray}
d\psi _{f}\left( x\right) =\sum\limits_{s}U\left( x,s\right) d\,\widetilde{ 
\psi }_{f}\left( s\right) \quad ,\quad d\psi _{f}^{\dag }\left( x\right)
=\sum\limits_{s}d\,\widetilde{\psi }_{f}^{\ast }\left( s\right) U^{\dag
}\left( s,x\right) 
 \end{eqnarray}which gives 
 \begin{eqnarray}
J=\det \left( U\right) \det \left( U^{\dag }\right) =\det \left( UU^{\dag
}\right) =1\;.
 \end{eqnarray}
Now let 
 \begin{eqnarray}
\widetilde{Z}_{f}=\int \mathcal{D}\,\widetilde{\psi }_{f}^{\ast }\left(
s\right) \,\mathcal{D}\,\widetilde{\psi }_{f}\left( s\right) \,\,e^{iS_{f}}
 \end{eqnarray}with the notation in this context now meaning that 
\begin{equation}
\widetilde{Z}_{f}=\prod_{s}\widetilde{z}_{f}\left( s\right) 
\end{equation} 
where 
\begin{eqnarray}
\widetilde{z}_{f}\left( s\right)  &=&i\int d\,\widetilde{\psi }_{f}^{\ast
}\left( s\right) \int \,d\,\widetilde{\psi }_{f}\left( s\right) \,e^{i\, 
\widetilde{\psi }_{f}^{\ast }\,\left( s\right) \,a\left( s\right) \, 
\widetilde{\psi }_{f}\left( s\right) } \\
&=&a\left( s\right) 
\end{eqnarray} 
so that 
 \begin{eqnarray}
Z_{f}=\widetilde{Z}_{f} \; .
 \end{eqnarray}This is the path integral for an arbitrary time interval (with the fields,
operator, and meaning of time left unchanged), so the Lorentzian path
integral $\widetilde{Z}_{f}$ will give the same results as the Euclidean
path integral $Z_{f}$ for any physical process. The same is true of more
general path integrals derived from more general operators, as long as they
can be put into Gaussian form.

When the inverse transformation from $\widetilde{\psi }_{f}$ to $\psi _{f}$
is performed, we obtain 
\begin{equation}
Z_{f}=\int \mathcal{D}\,\psi _{f}^{\dag }\left( x\right) \,\mathcal{D}
\,\,\psi _{f}\left( x\right) \,e^{iS_{f}}
\end{equation}
with $S_{f}$ having its form (\ref{eq5}) in the coordinate representation.

\section{Transformation to Lorentzian path integral: bosons}

For bosons we\ can again perform the transformation (\ref{eq7}) to obtain 
\begin{equation}
S_{b}=\sum\nolimits_{s}\widetilde{\psi }_{b}^{\ast }\,\left( s\right)
\,a\left( s\right) \,\widetilde{\psi }_{b}\left( s\right) \; .
\end{equation} 
The formal expression for the Euclidean path integral is 
\begin{eqnarray}
Z_{b} &=&\int \mathcal{D}\,\psi _{b}^{\dag }\left( x\right) \,\mathcal{D} 
\,\psi _{b}\left( x\right) \,\,\,e^{-S_{b}} \\
&=&\prod_{x,ra}\int_{-\infty }^{\,\infty }d\,\left( \mathrm{Re}\,\psi
_{b,ra}\left( x\right) \right) \,\int_{-\infty }^{\,\infty }d\,\left( 
\mathrm{Im}\,\psi _{b,ra}\left( x\right) \right) \,e^{-S_{b}} \\
&=&\prod_{s}z_{b}\left( s\right) 
\end{eqnarray} 
with 
\begin{eqnarray}
z_{b}\left( s\right)  &=&\int_{-\infty }^{\,\infty }d(\mathrm{Re}\, 
\widetilde{\psi }_{b}\left( s\right) )\int_{-\infty }^{\,\infty }d(\mathrm{Im 
}\,\widetilde{\psi }_{b}\left( s\right) )\,e^{-S_{b}} \, .
\end{eqnarray}

We will now show that this action can be put into a form which corresponds
to scalar bosonic fields plus their auxiliary fields. First, if the gauge
potentials $A_{\mu }^{i}$ were zero, we would have 
\begin{equation}
\,\,i\sigma ^{\mu }\partial _{\mu }U_{0}\left( x,s\right) =a_{0}\left(
s\right) U_{0}\left( x,s\right) \;.
\end{equation}
Then 
\begin{equation}
\,U_{0}\left( x,s\right) =\mathcal{V}^{-1/2}u\left( s\right) e^{ip_{s}\cdot
x}\;\;,\;\;p_{s}\cdot x=\eta _{\mu \nu }p_{s}^{\mu }x^{\nu }\;\;,\;\;\eta
_{\mu \nu }=diag\left( -1,1,1,1\right)
\end{equation}
(with $\mathcal{V}$ a four-dimensional normalization volume) gives 
\begin{equation}
-\eta _{\mu \nu }\sigma ^{\mu }p_{s}^{\nu }U_{0}\left( x,s\right)
=a_{0}\left( s\right) U_{0}\left( x,s\right)
\end{equation}
where $\sigma ^{\mu }$ implicitly multiplies the identity matrix for the
multicomponent function $U_{0}\left( x,s\right) $. A given 2-component
spinor $u_{r}\left( s\right) $ has two eigenstates of $p_{s}^{k}\sigma ^{k}$
: 
\begin{equation}
p_{s}^{k}\sigma ^{k}u_{r}^{+}\left( s\right) =\left\vert \overrightarrow{p}
_{s}\right\vert \sigma ^{k}u_{r}^{+}\left( s\right) \quad ,\quad
p_{s}^{k}\sigma ^{k}u_{r}^{-}\left( s\right) =-\left\vert \overrightarrow{p}
_{s}\right\vert u_{r}^{-}\left( s\right)
\end{equation}
where $\overrightarrow{p}_{s}$ is the 3-momentum and $\left\vert 
\overrightarrow{p}_{s}\right\vert =\left( p_{s}^{k}p_{s}^{k}\right) ^{1/2}$.
The multicomponent eigenstates of $i\sigma ^{\mu }\partial _{\mu }$ and
their eigenvalues $a_{0}\left( s\right) =p_{s}^{0}\mp $ $\left\vert 
\overrightarrow{p}_{s}\right\vert $ thus come in pairs, corresponding to
opposite helicities.

For nonzero $A_{\mu }^{i}$, the eigenvalues $a\left( s\right) $ will also
come in pairs, with one growing out of $a_{0}\left( s\right) $ and the other
out of its partner $a_{0}\left( s^{\prime }\right) $ as the $A_{\mu }^{i}$
are turned on. To see this, first write (\ref{eq8}) as 
\begin{equation}
\left( i\partial _{0}+A_{0}^{i}t_{i}\right) U\left( x,s\right) +\sigma
^{k}\left( i\partial _{k}+A_{k}^{i}t_{i}\right) U\left( x,s\right) =a\left(
s\right) U\left( x,s\right)   \label{eq36}
\end{equation} 
or 
\begin{equation}
\left( i\partial _{0}+A_{0}^{i}t_{i}\right) _{rr^{\prime }}U_{r^{\prime
}}\left( x,s\right) +P_{rr^{\prime }}U_{r^{\prime }}\left( x,s\right)
-a\left( s\right) \delta _{rr^{\prime }}U_{r^{\prime }}\left( x,s\right) =0
\label{eq37}
\end{equation} 
\begin{equation}
P_{rr^{\prime }}\equiv \sigma ^{k}\left( i\partial
_{k}+A_{k}^{i}t_{i}\right) _{rr^{\prime }}
\end{equation} 
with the usual implied summations over repeated indices. At fixed $r$,$ 
r^{\prime }$ (and $x,s$), apply a unitary matrix $u$ which will diagonalize
the $2\times 2$ matrix $P_{rr^{\prime }}$, bringing it into the form $ 
p_{rr^{\prime }}\sigma ^{3}+\overline{p}_{rr^{\prime }}\sigma ^{0}$, where $ 
p_{rr^{\prime }}$ and $\overline{p}_{rr^{\prime }}$ are 1-component
operators, while at the same time rotating the 2-component spinor $ 
U_{r^{\prime }}$: 
\begin{eqnarray}
uP_{rr^{\prime }}u^{\dag } &=&P_{rr^{\prime }}^{\prime }=p_{rr^{\prime
}}\sigma ^{3}+\overline{p}_{rr^{\prime }}\sigma ^{0}\quad ,\quad
U_{r^{\prime }}^{\prime }=uU_{r^{\prime }}\quad ,\quad uu^{\dag }=1 \\
\sigma ^{0} &=&\left( 
\begin{array}{cc}
1 & 0 \\ 
0 & 1 
\end{array} 
\right) \quad ,\quad \sigma ^{3}=\left( 
\begin{array}{cc}
1 & 0 \\ 
0 & -1 
\end{array} 
\right) .
\end{eqnarray} 
But $P_{rr^{\prime }}$ is traceless, and the trace is invariant under a
unitary transformation, so $\overline{p}_{rr^{\prime }}=0$. Then the second
term in (\ref{eq37}) becomes 
$u^{\dag }p_{rr^{\prime }}\sigma ^{3}U_{r^{\prime }}^{\prime }\left(
x,s\right) $. The two independent choices 
\begin{eqnarray}
U_{r^{\prime }}^{\prime }\left( x,s\right)  &\propto&\left( 
\begin{array}{c}
1 \\ 
0 
\end{array} 
\right) \quad ,\quad \sigma ^{3}U_{r^{\prime }}^{\prime }\left( x,s\right)
=U_{r^{\prime }}^{\prime }\left( x,s\right)  \\
U_{r^{\prime }}^{\prime }\left( x,s\right)  &\propto&\left( 
\begin{array}{c}
0 \\ 
1 
\end{array} 
\right) \quad ,\quad \sigma ^{3}U_{r^{\prime }}^{\prime }\left( x,s\right)
=-U_{r^{\prime }}^{\prime }\left( x,s\right) 
\end{eqnarray} 
give 
$\pm u^{\dag }p_{rr^{\prime }}U_{r^{\prime }}^{\prime }\left( x,s\right)$.
Now use $u^{\dag }U_{r^{\prime }}^{\prime }=U_{r^{\prime }}$ to obtain for 
(\ref{eq37})  
\begin{equation}
\left( i\partial _{0}+A_{0}^{i}t_{i}\right) _{rr^{\prime }}U_{r^{\prime
}}\left( x,s\right) \pm p_{rr^{\prime }}U_{r^{\prime }}\left( x,s\right)
-a\left( s\right) \delta _{rr^{\prime }}U_{r^{\prime }}\left( x,s\right) =0
\label{eq45}
\end{equation} 
so (\ref{eq36}) reduces to two $N\times N$ eigenvalue equations with
solutions 
\begin{eqnarray}
\left( i\partial _{0}+A_{0}^{i}t_{i}\right) U\left( x,s\right) +\sigma
^{k}\left( i\partial _{k}+A_{k}^{i}t_{i}\right) U\left( x,s\right) 
&=&a\left( s\right) U\left( x,s\right)   \label{eq27} \\
a\left( s\right)  &=&a_{1}\left( s\right) +a_{2}\left( s\right) 
\end{eqnarray} 
\begin{eqnarray}
\left( i\partial _{0}+A_{0}^{i}t_{i}\right) U\left( x,s^{\prime }\right)
+\sigma ^{k}\left( i\partial _{k}+A_{k}^{i}t_{i}\right) U\left( x,s^{\prime
}\right)  &=&a\left( s^{\prime }\right) U\left( x,s^{\prime }\right)  \\
a\left( s^{\prime }\right)  &=&a_{1}\left( s\right) -a_{2}\left( s\right) 
\label{eq30}
\end{eqnarray} 
where these equations define $a_{1}\left( s\right) $ and $a_{2}\left(
s\right) $. Notice that letting $\sigma ^{k}\rightarrow -\sigma ^{k}$ in 
(\ref{eq36}) reverses the signs in (\ref{eq45}), and changes the eigenvalue
of $U\left( x,s\right) $ to $a\left( s^{\prime }\right) =a_{1}\left(
s\right) -a_{2}\left( s\right) $.

The action for a single eigenvalue 
$\,a\left( s\right) $ and its partner $a\left( s^{\prime }\right) $ is 
\begin{eqnarray}
\widetilde{s}_{b}\left( s\right) &=&\widetilde{\psi }_{b}^{\ast }\left(
s\right) \,a\left( s\right) \,\widetilde{\psi }_{b}\left( s\right) + 
\widetilde{\psi }_{b}^{\ast }\left( s^{\prime }\right) \,a\left( s^{\prime
}\right) \,\widetilde{\psi }_{b}\left( s^{\prime }\right) \  \\
&=&\widetilde{\psi }_{b}^{\ast }\left( s\right) \left( a_{1}\left( s\right)
+a_{2}\left( s\right) \right) \widetilde{\psi }_{b}\left( s\right) + 
\widetilde{\psi }_{b}^{\ast }\left( s^{\prime }\right) \left( a_{1}\left(
s\right) -a_{2}\left( s\right) \right) \widetilde{\psi }_{b}\left( s^{\prime
}\right) \;.
\end{eqnarray}
There are 4 cases: For $a_{1}\left( s\right) >0$ and $a_{2}\left( s\right) >0
$, let 
\begin{eqnarray}
\widetilde{\psi }_{b}\left( s^{\prime }\right)  &=&a\left( s\right) ^{1/2} 
\widetilde{\phi }_{b}\left( s\right) =\left( a_{1}\left( s\right)
+a_{2}\left( s\right) \right) ^{1/2}\widetilde{\phi }_{b}\left( s\right) 
\label{eq16} \\
\widetilde{\psi }_{b}\left( s\right)  &=&a\left( s\right) ^{-1/2}\widetilde{F 
}_{b}\left( s\right) =\left( a_{1}\left( s\right) +a_{2}\left( s\right)
\right) ^{-1/2}\widetilde{F}_{b}\left( s \right) 
\end{eqnarray} 
and for $a_{1}\left( s\right) >0$ and $a_{2}\left( s\right) <0$ 
\begin{eqnarray}
\widetilde{\psi }_{b}\left( s\right)  &=&a\left( s^{\prime }\right) ^{1/2} 
\widetilde{\phi }_{b}\left( s\right) =\left( a_{1}\left( s\right)
-a_{2}\left( s\right) \right) ^{1/2}\widetilde{\phi }_{b}\left( s \right) \\
\widetilde{\psi }_{b}\left( s^{\prime }\right)  &=&a\left( s^{\prime
}\right) ^{-1/2}\widetilde{F}_{b}\left( s \right) =\left(
a_{1}\left( s\right) -a_{2}\left( s\right) \right) ^{-1/2}\widetilde{F} 
_{b}\left( s \right)   \label{eq19}
\end{eqnarray} 
so that for both of these first two cases 
 \begin{eqnarray}
\widetilde{s}_{b}\left( s\right) =\widetilde{\phi }_{b}^{\ast }\left(
s\right) \widetilde{a}\left( s\right) \widetilde{\phi }_{b}\left( s\right) + 
\widetilde{F}_{b}^{\ast }\left( s\right) \widetilde{F}_{b}\left( s\right)
\quad ,\quad a_{1}\left( s\right) >0
 \end{eqnarray}where 
 \begin{eqnarray}
\widetilde{a}\left( s\right) =a\left( s\right) a\left( s^{\prime }\right)
=a_{1}\left( s\right) ^{2}-a_{2}\left( s\right) ^{2}.
 \end{eqnarray}For $a_{1}\left( s\right) <0$ and $a_{2}\left( s\right) <0$, let 
\begin{eqnarray}
\widetilde{\psi }_{b}\left( s^{\prime }\right)  &=&\left( -a\left( s\right)
\right) ^{1/2}\widetilde{\phi }_{b}\left( s\right) =\left( -a_{1}\left(
s\right) -a_{2}\left( s\right) \right) ^{1/2}\widetilde{\phi }_{b}\left(
s\right)   \label{eq22} \\
\widetilde{\psi }_{b}\left( s\right)  &=&\left( -a\left( s\right) \right)
^{-1/2}\widetilde{F}_{b}\left( s\right) =\left( -a_{1}\left( s\right)
-a_{2}\left( s\right) \right) ^{-1/2}\widetilde{F}_{b}\left( s\right) 
\end{eqnarray} 
and for $a_{1}\left( s\right) <0$ and $a_{2}\left( s\right) >0$ 
\begin{eqnarray}
\widetilde{\psi }_{b}\left( s\right)  &=&\left( -a\left( s^{\prime }\right)
\right) ^{1/2}\widetilde{\phi }_{b}\left( s\right) =\left( -a_{1}\left(
s\right) +a_{2}\left( s\right) \right) ^{1/2}\widetilde{\phi }_{b}\left(
s\right)  \\
\widetilde{\psi }_{b}\left( s^{\prime }\right)  &=&\left( -a\left( s^{\prime
}\right) \right) ^{-1/2}\widetilde{F}_{b}\left( s\right) =\left(
-a_{1}\left( s\right) +a_{2}\left( s\right) \right) ^{-1/2}\widetilde{F} 
_{b}\left( s\right)   \label{eq25}
\end{eqnarray} 
so for each of these last two cases 
\begin{equation}
\widetilde{s}_{b}\left( s\right) =-\left[ \widetilde{\phi }_{b}^{\ast
}\left( s\right) \widetilde{a}\left( s\right) \widetilde{\phi }_{b}\left(
s\right) +\widetilde{F}_{b}^{\ast }\left( s\right) \widetilde{F}_{b}\left(
s\right) \right] \quad ,\quad a_{1}\left( s\right) <0 \; .
\end{equation} 
Then we have 
\begin{eqnarray}
S_{b} &=&\sum\nolimits_{s}^{^{\prime }}\widetilde{s}_{b}\left( s\right)  \\
&=&\sum\limits_{a_{1}\left( s\right) >0}^{\prime }\left[ \widetilde{\phi } 
_{b}^{\ast }\left( s\right) \widetilde{a}\left( s\right) \widetilde{\phi } 
_{b}\left( s\right) +\widetilde{F}_{b}^{\ast }\left( s\right) \widetilde{F} 
_{b}\left( s\right) \right]   \nonumber \\
&&\hspace{2cm}-\sum\limits_{a_{1}\left( s\right) <0}^{\prime }\left[ 
\widetilde{\phi }_{b}^{\ast }\left( s\right) \widetilde{a}\left( s\right) 
\widetilde{\phi }_{b}\left( s\right) +\widetilde{F}_{b}^{\ast }\left(
s\right) \widetilde{F}_{b}\left( s\right) \right]   \nonumber
\end{eqnarray} 
where a prime on a summation or product over $s$ means that only one member
of an $s,s^{\prime }$ pair (as defined in (\ref{eq27})-(\ref{eq30})) is
included, so that there are only $N$ terms rather than $2N$.

All of the transformations above from $\widetilde{\psi }_{b}$ to $\widetilde{ 
\phi }_{b}$ and $\widetilde{F}_{b}$ have the form 
\begin{equation}
\widetilde{\psi }_{b}\left( s_{1}\right) =A\left( s\right) ^{1/2}\widetilde{ 
\phi }_{b}\left( s\right) \quad ,\quad \widetilde{\psi }_{b}\left(
s_{2}\right) =A\left( s\right) ^{-1/2}\widetilde{F}_{b}\left( s\right) 
\end{equation} 
so that 
\begin{equation}
d\widetilde{\psi }_{b}\left( s_{1}\right) =A\left( s\right) ^{1/2}d 
\widetilde{\phi }_{b}\left( s\right) \quad ,\quad d\widetilde{\psi } 
_{b}\left( s_{2}\right) =A\left( s\right) ^{-1/2}d\widetilde{F}_{b}\left(
s\right) 
\end{equation} 
and the Jacobian is 
 \begin{eqnarray}
J^{\prime }=\prod\nolimits_{s}A\left( s\right) ^{1/2}A\left( s\right)
^{-1/2}=1 \; .
 \end{eqnarray}These transformations lead to the formal result 
\begin{equation}
Z_{b}=\prod_{a_{1}\left( s\right) >0}^{\prime }z_{b}\left( s\right) \cdot
\prod_{a_{1}\left( s\right) <0}^{\prime }z_{b}\left( s\right)   \label{eq47}
\end{equation} 
\begin{eqnarray}
z_{b}\left( s\right)  &=&\int_{-\infty }^{\,\infty }d(\mathrm{Re}\, 
\widetilde{\phi }_{b}\left( s\right) )\int_{-\infty }^{\,\infty }d(\mathrm{Im 
}\,\widetilde{\phi }_{b}\left( s\right) )\int_{-\infty }^{\,\infty }d( 
\mathrm{Re}\,\widetilde{F}_{b}\left( s\right) )\int_{-\infty }^{\,\infty }d( 
\mathrm{Im}\,\widetilde{F}_{b}\left( s\right) )  \nonumber \\
\hspace{-0.4cm} &\times &\,e^{-\widetilde{a}\left( s\right) \left[ \left( 
\mathrm{Re}\,\widetilde{\phi }_{b}\left( s\right) \right) ^{2}+\left( 
\mathrm{Im}\,\widetilde{\phi }_{b}\left( s\right) \right) ^{2}\right] }e^{- 
\left[ \left( \mathrm{Re}\,\widetilde{F}_{b}\left( s\right) \right)
^{2}+\left( \mathrm{Im}\,\widetilde{F}_{b}\left( s\right) \right) ^{2}\right]
}\;,\;a_{1}\left( s\right) >0
\end{eqnarray} 
\begin{eqnarray}
z_{b}\left( s\right)  &=&\int_{-\infty }^{\,\infty }d(\mathrm{Re}\, 
\widetilde{\phi }_{b}\left( s\right) )\int_{-\infty }^{\,\infty }d(\mathrm{Im 
}\,\widetilde{\phi }_{b}\left( s\right) )\int_{-\infty }^{\,\infty }d( 
\mathrm{Re}\,\widetilde{F}_{b}\left( s\right) )\int_{-\infty }^{\,\infty }d( 
\mathrm{Im}\,\widetilde{F}_{b}\left( s\right) )  \nonumber \\
\hspace{-0.4cm} &\times &\,e^{\widetilde{a}\left( s\right) \left[ \left( 
\mathrm{Re}\,\widetilde{\phi }_{b}\left( s\right) \right) ^{2}+\left( 
\mathrm{Im}\,\widetilde{\phi }_{b}\left( s\right) \right) ^{2}\right] }e^{ 
\left[ \left( \mathrm{Re}\,\widetilde{F}_{b}\left( s\right) \right)
^{2}+\left( \mathrm{Im}\,\widetilde{F}_{b}\left( s\right) \right) ^{2}\right]
}\;,\;a_{1}\left( s\right) <0 \; .
\end{eqnarray}

At this point we encounter a difficulty which is not present for fermions,
since the integral over Grassmann variables is well-defined for both
positive and negative $a\left( s\right) $, whereas the corresponding
integrals above, over ordinary commuting variables, are divergent for the
states with either 
$\widetilde{a}\left( s\right) <0$ or $a_{1}\left( s\right) <0$. This
divergence results from the approximate linearization that led to (\ref{eq3}), 
and will ultimately be controlled by various nonlinear effects, beginning
with a   self-interaction term involving $\left( \psi _{b}^{\dag }\,\left(
x\right) \psi _{b}\left( x\right) \right) ^{2}$ which is present in the
original theory, but also including gauge interactions and various other
complications which certainly lie beyond the simple treatment given here. We
will therefore omit these states, which have a different status and require
special treatment, in the expansions of $\phi _{b}\left( x\right) $ and $ 
F_{b}\left( x\right) $: 
\begin{eqnarray}
\phi _{b}\left( x\right)  &=&\sum\limits_{s>0}\widetilde{U}\left( x,s\right)
\,\widetilde{\phi }_{b}\left( s\right) \quad ,\quad \widetilde{\phi } 
_{b}\left( s\right) =\int d^{4}x\,\widetilde{U}^{\ast }\left( s,x\right)
\,\phi _{b}\left( x\right)   \label{eq48} \\
F_{b}\left( x\right)  &=&\sum\limits_{s>0}\widetilde{U}\left( x,s\right) \, 
\widetilde{F}_{b}\left( s\right) \quad ,\quad \,\widetilde{F}_{b}\left(
s\right) =\int d^{4}x\,\widetilde{U}^{\ast }\left( s,x\right) \,F_{b}\left(
x\right)   \label{eq49}
\end{eqnarray} 
where $s>0$ means that $a_{1}\left( s\right) >0$ and $\widetilde{a}\left(
s\right) >0$. 

Here $\widetilde{U}$ is an $N\times N$ matrix (since there is no longer a
spinor index $a$, and the number of values of $s$ has also been reduced by a
factor of $2$) which satisfies 
\begin{equation}
\,\,\eta ^{\mu \nu }D_{\mu }D_{\nu }\widetilde{U}\left( x,s\right) =\left[
a_{1}\left( s\right) ^{2}-a_{2}\left( s\right) ^{2}\right] U\left(
x,s\right) =\widetilde{a}\left( s\right) U\left( x,s\right)   \label{eq75}
\end{equation} 
\begin{equation}
\int d^{4}x\,\widetilde{U}^{\ast }\left( s,x\right) \widetilde{U}\left(
x,s^{\prime }\right) \,=\delta _{ss^{\prime }} \; .
\end{equation} 
I.e., we assume the existence of basis functions $\widetilde{U}\left(
x,s\right) $ which are the appropriate solutions of (\ref{eq75}). 
In cases where an appropriate set 
of such basis functions does not exist, these bosons will exhibit further
nonstandard behavior.

In the case of free fields (i.e. with $A_{\mu }^{i}=0$), we
have 
\begin{eqnarray}
\widetilde{U}\left( x,s\right)  &=& \mathcal{V}^{-1/2}e^{ip_{s}\cdot x}
\\ 
a_{1}\left( s\right)  &=& \omega \equiv p_{s}^{0}\quad ,\quad a_{2}\left(
s\right) = \pm \left\vert \overrightarrow{p}_{s}\right\vert \quad ,\quad
\left\vert \overrightarrow{p}_{s}\right\vert \equiv \left(
p_{s}^{k}p_{s}^{k}\right) ^{1/2} \\
\widetilde{a}\left( s\right) &=& \left( p_{s}^{0}\pm \left\vert 
\overrightarrow{p}_{s}\right\vert \right) \left( p_{s}^{0}\mp \left\vert 
\overrightarrow{p}_{s}\right\vert \right) =\left( p_{s}^{0}\right)
^{2}-\left\vert \overrightarrow{p}_{s}\right\vert ^{2} 
\end{eqnarray}
and
\begin{eqnarray}
\eta ^{\mu \nu }D_{\mu }D_{\nu }\widetilde{U}\left( x,s\right) = \eta ^{\mu
\nu }\partial _{\mu }\partial _{\nu }\widetilde{U}\left( x,s\right) =\left[
\left( p_{s}^{0}\right) ^{2}-\left\vert \overrightarrow{p}_{s}\right\vert
^{2}\right] \widetilde{U}\left( x,s\right) \; .
\end{eqnarray}
Also, $s>0$ then means that $\omega >0$ and 
\begin{eqnarray}
\omega >\left\vert \overrightarrow{p}_{s}\right\vert \, .
\end{eqnarray}

With a return to the general case, (\ref{eq47}) becomes 
\begin{equation}
Z_{b}=\prod_{s>0}^{\prime }z_{b}\left( s\right) 
\end{equation}
where 
\begin{eqnarray}
z_{b}\left( s \right) =
\frac{\pi }{\widetilde{a}\left( s\right) }\cdot \frac{\pi }{1}\quad \text{
for } s>0 \; .
\end{eqnarray}

Now let 
 \begin{eqnarray}
\widetilde{Z}_{b} &=& \int \mathcal{D}\,\widetilde{\phi } 
_{b}^{\dag }\,\left( s\right) \mathcal{D}\,\widetilde{\phi }_{b}\left(
s\right) \,\mathcal{D}\,\widetilde{F}_{b}^{\dag }\,\left( s\right) \mathcal{D 
}\,\widetilde{F}_{b}\left( s\right) \,\,e^{iS_{b}} \\
 &\equiv&  \prod_{s>0}^{\prime }\widetilde{z}_{b}\left( s\right)
 \end{eqnarray}
with 
\begin{eqnarray}
\widetilde{z}_{b}\left( s\right)  &\equiv &-\int_{-\infty }^{\,\infty }d( 
\mathrm{Re}\,\widetilde{\phi }_{b}\left( s\right) )\int_{-\infty }^{\,\infty
}d(\mathrm{Im}\,\widetilde{\phi }_{b}\left( s\right) )\int_{-\infty
}^{\,\infty }d(\mathrm{Re}\,\widetilde{F}_{b}\left( s\right) )\int_{-\infty
}^{\,\infty }d(\mathrm{Im}\,\widetilde{F}_{b}\left( s\right) )  \nonumber \\
&&\hspace{1cm}\times \,e^{i\widetilde{a}\left( s\right) \left[ \left( 
\mathrm{Re}\,\widetilde{\phi }_{b}\left( s\right) \right) ^{2}+\left( 
\mathrm{Im}\,\widetilde{\phi }_{b}\left( s\right) \right) ^{2}\right] }e^{i 
\left[ \left( \mathrm{Re}\,\widetilde{F}_{b}\left( s\right) \right)
^{2}+\left( \mathrm{Im}\,\widetilde{F}_{b}\left( s\right) \right) ^{2}\right]
} \\
&=&\frac{\pi }{\widetilde{a}\left( s\right) }\cdot \frac{\pi }{1}\quad \text{
for } s>0
\end{eqnarray} 
since $\int_{-\infty }^{\,\infty }dx\,\int_{-\infty }^{\,\infty }dy\,\exp
\left( \,ia\left( x^{2}+y^{2}\right) \right) =i\pi /a $.
(Nuances of Lorentzian path integrals are discussed in, e.g., Peskin and
Schroeder\cite{peskin}.)

We have then obtained $Z_{b}=\widetilde{Z}_{b}$, or after a transformation
to the coordinate representation via (\ref{eq48}) and (\ref{eq49}), 
\begin{eqnarray}
Z_{b} &=&\int \mathcal{D}\,\phi _{b}^{\dag }\,\left( x\right) \mathcal{D} 
\,\phi _{b}\left( x\right) \,\mathcal{D}\,F_{b}^{\dag }\,\left( x\right) 
\mathcal{D}\,F_{b}\left( x\right) \,\,e^{iS_{b}} \\
S_{b} &=&\int d^{4}x\,\left[ \phi _{b}^{\dag }\left( x\right) \,\eta ^{\mu
\nu }D_{\mu }D_{\nu }\phi _{b}\left( x\right) +F_{b}^{\dag }\left( x\right)
F_{b}\left( x\right) \right]  \label{eq100} .
\end{eqnarray} 
Again, this is the path integral for an arbitrary time interval, so the
Lorentzian path integral $\widetilde{Z}_{b}$ will give the same results as
the Euclidean path integral $Z_{b}$ for any physical process, and the same
is true for more general path integrals derived from more general operators.

Recall, however, that the states with $s<0$ have been omitted from the
expansion of $\phi _{b}\left( x\right) $ and $F_{b}\left( x\right) $, so
these bosonic fields should exhibit nonstandard behavior, and this
feature may provide the most testable new prediction of the present theory.

\section{Supersymmetry}

The total action for fermions and bosons is 
\begin{eqnarray}
\hspace{-0.2cm}S &=&S_{f}+S_{b} \\
&=&\int d^{4}x\,\left[ \psi _{f}^{\dag }\left( x\right) \,i\sigma ^{\mu
}D_{\mu }\,\psi _{f}\left( x\right) +\phi _{b}^{\dag }\left( x\right) \eta
^{\mu \nu }D_{\mu }D_{v}\phi _{b}\left( x\right) +F_{b}^{\dag }\left(
x\right) F_{b}\left( x\right) \right] 
\end{eqnarray} 
which in a general coordinate system becomes 
\begin{equation}
S=\int d^{4}x\,e\,\left[ \psi ^{\dag }\left( x\right) \,ie_{\alpha }^{\mu
}\,\sigma ^{\alpha }\widetilde{D}_{\mu }\,\psi \left( x\right) -g^{\mu \nu
}\left( \widetilde{D}_{\mu }\phi \left( x\right) \right) ^{\dag }\widetilde{D 
}_{v}\phi \left( x\right) +F^{\dag }\left( x\right) F\left( x\right) \right] 
\label{eq62}
\end{equation} 
where $g_{\mu \nu }$ is the metric tensor,
$e =\det e_{\mu }^{\alpha }=\left( -\det g_{\mu \nu }\right) ^{1/2} $,
$\widetilde{D}_{\mu }=D_{\mu }+e^{-1/2}\partial _{\mu }e^{1/2} $, and 
\begin{eqnarray}
\psi \left( x\right) =e^{-1/2}\;\psi _{f}\left( x\right) \quad ,\quad \phi
\left( x\right) =e^{-1/2}\;\phi _{b}\left( x\right) \quad ,\quad F\left(
x\right) =e^{-1/2}F_{b}\left( x\right) .
 \end{eqnarray}
We thus obtain the standard basic form for a supersymmetric action, where the
fields $\phi $, $F$, and $\psi $ respectively consist of 1-component complex
scalar bosonic fields$,$ 1-component complex scalar auxiliary fields, and
2-component spin 1/2 fermionic fields $\psi $. These fields span the various
physical representations of the fundamental gauge group, which must be $SO(N)
$ (e.g., $SO(10)$) in the present theory. I.e., $\psi $ includes all the
Standard Model fermions and the Higgsinos, and $\phi $ includes the
sfermions and Higgses.

\section{Higher-derivative terms in the initial bosonic action}

It was mentioned below Eq. (3.21) in Ref.~2 that higher-derivative terms are
required in the initial bosonic action in order for the action in the
internal space to be finite. It is easy to revise the treatment in Ref.~3
between Eqs. (73) and (87) to obtain the lowest-order such term. First (for
better-defined statistical counting) we choose the length scale $a$ in
external space to be the same as the original fundamental length scale $a_{0}
$ and rewrite Eq. (73) of Ref.~3 as 
 \begin{eqnarray}
\overline{S}=S_{0}+\sum_{\overline{x},k}a\left\langle \Delta
\rho _{k}\right\rangle a_{0}^{D}-\sum_{\overline{x},k}b\left[ \left\langle
\Delta \rho _{k}\right\rangle ^{2}+\left\langle \left( \delta \rho
_{k}\right) ^{2}\right\rangle \right] \left( a_{0}^{D}\right) ^{2}\;.
 \end{eqnarray}We then retain the second-order term in $\delta \rho _{k}$: 
\begin{equation}
\delta \rho _{k}=\frac{\partial \Delta \rho _{k}}{\partial \overline{x}^{M}} 
\delta x+\frac{1}{2}\frac{\partial ^{2}\Delta \rho _{k}}{\partial \left( 
\overline{x}^{M}\right) ^{2}}\left( \delta x\right) ^{2}\;.
\end{equation} 
With $\partial \Delta \rho _{k}/\partial \overline{x}^{M}=\partial \left(
\rho _{k}-\overline{\rho }\right) /\partial \overline{x}^{M}=\partial \rho
_{k}/\partial \overline{x}^{M}$, it follows that 
\begin{eqnarray}
\left\langle \left( \delta \rho _{k}\right) ^{2}\right\rangle  &=&\sum_{M}
\left[ \left( \frac{\partial \rho _{k}}{\partial \overline{x}^{M}}\right)
^{2}\left( \frac{a_{0}}{2}\right) ^{2}+\left( \frac{1}{2}\frac{\partial
^{2}\rho _{k}}{\partial \left( \overline{x}^{M}\right) ^{2}}\right)
^{2}\left( \frac{a_{0}}{2}\right) ^{4}\right]  \\
&=&\sum_{M}\rho _{k}\,a_{0}^{2}\left[ \left( \frac{\partial \phi _{k}}{ 
\partial \overline{x}^{M}}\right) ^{2}+\frac{a_{0}^{2}}{16}\left( \frac{ 
\partial ^{2}\phi _{k}}{\partial \left( \overline{x}^{M}\right) ^{2}}\right)
^{2}\right] \;
\end{eqnarray} 
with the higher-order term involving the first derivative neglected. In the
continuum limit, $\sum_{\overline{x}}a_{0}^{D}\rightarrow
\int_{a_{0}}^{\infty }d^{D}x$, this leads to 
\begin{equation}
\overline{S}=S_{0}^{\prime }+\int_{a_{0}}^{\infty
}d^{D}x\,\,\sum_{k}\left\{ \frac{\mu }{m}\phi _{k}^{2}-\frac{1}{2m^{2}} 
\sum_{M}\left[ \left( \frac{\partial \phi _{k}}{\partial x^{M}}\right) ^{2}+ 
\frac{a_{0}^{2}}{16}\left( \frac{\partial ^{2}\phi _{k}}{\partial \left(
x^{M}\right) ^{2}}\right) ^{2}\right] \right\} 
\end{equation} 
with the lower limit $a_{0}$ automatically providing an ultimate ultraviolet
cutoff. Eq. (87) of Ref.~3 is then replaced by 
 \begin{eqnarray}
\overline{S}_{b} = \int_{a_{0}}^{\infty }d^{D}x\,\bigg\{\frac{1}{ 
2m^{2}}\left[ \frac{\partial \Psi _{b}^{\dagger }}{\partial x^{M}}\frac{ 
\partial \Psi _{b}}{\partial x^{M}}+\frac{a_{0}^{2}}{16}\frac{\partial
^{2}\Psi _{b}^{\dagger }}{\partial \left( x^{M}\right) ^{2}}\frac{\partial
^{2}\Psi _{b}}{\partial \left( x^{M}\right) ^{2}}\right] \nonumber \\
-\mu \,\Psi _{b}^{\dagger }\Psi _{b}+i\widetilde{V}\,\Psi _{b}^{\dagger }
\Psi _{b}\bigg\}. 
 \end{eqnarray}
Ordinarily we can let $a_{0} \rightarrow 0$, but both the nonzero lower limit 
and the higher-derivative term in the action can be relevant in the
internal space, where the length scales can be comparable to $a_{0}$, which
may itself be regarded as comparable to the Planck length. Finally, we
emphasize that the randomly fluctuating imaginary potential $i\widetilde{V}$
is a separate postulate of the theory. As mentioned below, the present
theory is based on both statistical counting and these stochastic
fluctuations, as well as the specific symmetry-breaking or ``geography'' of
our universe.

\section{Gravity and cosmological constant}

According to (\ref{eq62}), the coupling of matter to gravity is very
nearly the same as in standard general relativity. However, if $S$ is
written in terms of the original fields $\psi _{f}$ and $\phi _{b}$, there
is no factor of $e$. In other words, in the present theory the original
action has the form 
 \begin{eqnarray}
S=\int d^{4}x\,\mathcal{L}
 \end{eqnarray}whereas in standard physics it has the form 
 \begin{eqnarray}
S=\int d^{4}x\,e\,\overline{\mathcal{L}}\;.
 \end{eqnarray}For an $\mathcal{L}$ corresponding to a fixed vacuum energy density, there
is then no coupling to gravity in the present theory, and the usual
cosmological constant vanishes. This point was already made in Ref.~1, where
the \textquotedblleft cosmological constant\textquotedblright\ was defined
to be the usual contribution to the stress-energy tensor from a constant
vacuum Lagrangian density $\mathcal{L}_{0}$, which results from the factor
of $e$. However, as was also pointed out in this 1996 paper,
\textquotedblleft There may be a much weaker term involving $\delta \mathcal{ 
L}_{0}/\delta g^{\mu \nu }$, but this appears to be consistent with
observation.\textquotedblright 

This much weaker term we now interpret to be a ``diamagnetic response'' of
vacuum fields to changes in both the vierbein and gauge fields, which
results from a shifting of the energies of the vacuum states when fields are
applied, just as the energies of the electrons in a metal are shifted by the
application of a magnetic field. We postulate that this effect produces
contributions to the action which are consistent with the
general coordinate invariance and gauge symmetry of the present theory. The
lowest-order such contributions are, of course, the Maxwell-Yang-Mills and
Einstein-Hilbert actions, plus a relatively weak cosmological constant
arising from this same mechanism: 
\begin{eqnarray}
\mathcal{L}_{g} = -\frac{1}{4}g_{0}^{-2} e F_{\mu \nu }^{i}F_{\rho \sigma
}^{i}g^{\mu \rho }g^{\nu \sigma } \quad , \quad
\mathcal{L}_{G} = e\,\Lambda
+\ell _{P}^{-2}e~^{(4)}R \; .
\end{eqnarray}
Here $g_{0}$ is the coupling constant for the fundamental gauge group (e.g.$ 
SO(10)$), $\Lambda $ is a constant, and $\ell _{P}^{2}=16\pi G$. These terms
are analogous to the usual contributions to the free energy from Landau
diamagnetism in a metal.

The actions for gauginos and gravitinos are postulated to have a similar
origin, as the vacuum responds to these fields. Particle masses and Yukawa
couplings are postulated to arise from supersymmetry breaking and radiative
corrections.

As pointed out in Ref.~1, the above gauge and gravitational curvatures
require that the order parameter contain a superposition of configurations
with topological defects (without which there could be no curvature). Here
we do not attempt to discuss these defects in detail, but we now interpret
them as 1-dimensional defect lines in 4-dimensional spacetime, analogous to
vortex lines in a superfluid.

There is clearly a lot of work remaining to be done -- including actual
predictions for experiment -- but the theory is relatively close to
real-world physics, and the following arise as emergent properties from a
fundamental statistical picture: Lorentz invariance, the general form of
Standard-Model physics, an $SO(N)$ fundamental gauge theory (with e.g. $ 
SO(10)$ permitting coupling constant unification and neutrino masses),
supersymmetry, a gravitational metric with the form $\left( -,+,+,+\right) $
, the correct coupling of matter fields to gravity, vanishing of the usual
cosmological constant, and a mechanism for the origin of spacetime and
fields.

The new predictions of the present theory appear to be subtle, but include
Lorentz violation at very high energies and nonstandard behavior of scalar
bosons.

\section{Conclusion}

For a theory to be viable, it must be mathematically consistent, its
premises must lead to testable predictions, and these predictions must be
consistent with experiment and observation. The theory presented here
appears to satisfy these requirements, although it is still very far from
complete.

Experiment should soon confront theory with more stringent constraints. For
example, supersymmetry~\cite{kane}, fundamental scalar
bosons~\cite{gunion}, 
and $SO(N)$ grand unification~\cite{barger} seem to be unavoidable
consequences of the theory presented here, but there is as yet no direct
evidence for any of these extensions of established physics.

The present theory starts with a picture which is far from that envisioned
in more orthodox approaches: There are initially no laws, and instead all
possibilities are realized with equal probability. The observed laws of
Nature are emergent phenomena, which result from statistical counting and
stochastic fluctuations, together with the specific symmetry-breaking (or
``geographical features'') of our universe.

It is reassuring that such an unconventional picture ultimately leads back
to both established physics and standard extensions like the three mentioned
above. Perhaps this fact helps to demonstrate the robustness and naturalness
of these extensions, and the importance of experimental searches for
supersymmetric partners, dark matter, Higgs bosons, the various consequences
of grand unification, and related phenomena in cosmology and astrophysics.

The present theory shares several central concepts with string theory --
namely supersymmetry, higher dimensions, and topological defects -- perhaps
indicating that these elements may be inescapable in a truly fundamental
theory.

\section{Acknowledgements}

I have benefitted greatly from many discussions with Seiichirou Yokoo and
Zorawar Wadiasingh. In particular, Seiichirou Yokoo obtained a determinantal
transformation for free fields which was a precursor of the explicit
transformation of fields in Eqs. (\ref{eq16})-(\ref{eq19}) and 
(\ref{eq22})-(\ref{eq25}).

\end{document}